\documentclass[11pt, oneside]{article}   	
\usepackage{geometry}                		
\usepackage[parfill]{parskip}    		
\usepackage{graphicx}				
\usepackage{amssymb}
\usepackage{subfigure}
\usepackage{amsmath}
\usepackage{hyperref}
\usepackage{float} 
\usepackage{multirow} 
\usepackage{booktabs} 
\usepackage{authblk} 

\title{The Similarity of Global Value Chains: \\ A Network-Based Measure}

\author[1]{Zhen Zhu}
\author[1]{Greg Morrison}
\author[1]{Michelangelo Puliga}
\author[1]{Alessandro Chessa}
\author[1,2]{Massimo Riccaboni}

\affil[1]{IMT Institute for Advanced Studies, Lucca, Italy}
\affil[2]{DMSI, K. U. Leuven, Leuven, Belgium}

\date{\today}

\begin{document}

\maketitle

\begin{abstract}
\noindent International trade has been increasingly organized in the form of global value chains (GVCs) where different stages of production are located in different countries. This recent phenomenon has substantial consequences for both trade policy design at the national or regional level and business decision making at the firm level.  In this paper, we provide a new method for comparing GVCs across countries and over time. First, we use the World Input-Output Database (WIOD) to construct both the upstream and downstream global value networks, where the nodes are individual sectors in different countries and the links are the value-added contribution relationships.  Second, we introduce a network-based measure of node similarity to compare the GVCs between any pair of countries for each sector and each year available in the WIOD.  
Our network-based similarity is a better measure for node comparison than the existing ones because it takes into account all the direct and indirect relationships between country-sector pairs, is applicable to both directed and weighted networks with self-loops, and takes into account externally defined node attributes. 
As a result, our measure of similarity reveals the most intensive interactions among the GVCs across countries and over time. From 1995 to 2011, the average similarity between sectors and countries have clear increasing trends, which are temporarily interrupted by the recent economic crisis.
This measure of the similarity of GVCs provides quantitative answers to important questions about dependency, sustainability, risk, and competition
in the global production system. 
\end{abstract}

\section{Introduction}

International trade has been increasingly characterized by the content of intermediate inputs~\cite{johnson2012accounting,johnson2014five} and by the formation of {\em{global value chains}} (GVCs)~\cite{feenstra1999impact,hummels2001nature,grossman2008trading,costinot2013elementary,amador2014global,baldwin2014supply,los2014global}. Thanks to the development of transportation, information, and communications technologies, different stages of production can be allocated and coordinated across borders. For instance, merely 3\% of the total value-added of China's exports of iPhones and laptop computers in 2009 is sourced from China itself, while the remaining 97\% is from other countries such as the United States, Japan, and South Korea~\cite{xing2014china}.  

With {\em{global multi-regional input-output}} (GMRIO) tables becoming available~\cite{tukker2013global}, the phenomenon of GVCs has been explored extensively in recent years by both theoretical modeling~\cite{grossman2008trading,baldwin2013spiders,costinot2013elementary} and empirical measurements~\cite{feenstra1999impact,hummels2001nature,johnson2012accounting,baldwin2014supply,johnson2014five,los2014global,timmer2014slicing,zhu2015global}. 
Although previous studies can tell us how `global' the GVCs are by measuring the foreign value-added content of exports for a given sector or country, this approach simply ignores the interdependence and interconnectedness of the GVCs (as an exception, see ~\cite{zhu2015global} where the network structure of the GVCs at the sector level is taken into account and simplified by the tree topology). The notion of GVCs has been useful in capturing the fact that different stages of production are organized across multiple countries, but the global production sharing at micro level (e.g., for a certain product such as iPhone) can be performed in a wide range of configurations, including a chain (or ``snake''), star (or ``spider''), or any network topology in between~\cite{baldwin2013spiders}. More importantly, at the aggregated sector level the GVCs are necessarily embedded in a global production network, where significant value-added contributions flow between sectors located in different countries.  Any measure of the GVCs ignoring the network structure would incur a great loss of information, and so the GVCs can only be meaningfully compared if the network structure is accounted for. 

Our paper is also related to the longstanding literature on export similarity. Since the seminal work of Finger and Kreinin~\cite{finger1979}, multiple measures of similarity have been introduced in the empirical study of international trade to calculate the overlap between the distributions of exports or imports by commodity groups of two countries to the market of third countries~\cite{lloyd2004}.
However, traditional measures of export similarity do not take into account the fragmentation of global production, which accounts for two-thirds of international trade~\cite{johnson2012accounting}. 

To fill the gaps in the literature, we introduce a network-based measure of similarity between the GVCs, which may provide possible insights into node clustering or community detection~\cite{MorrisonPLOS12,GirvanPNAS02,ZhouPRE03}, link prediction~\cite{LuPhysicaA11,LuPRE09,ZhouEPJB09}, and block modeling~\cite{lerner2005role,ReichardtEJPB07,GuimeraJStatMech05}.  Decades of literature has implemented measures of {\em{structural equivalence}} between nodes, with equivalent nodes strongly connected to the same neighbors~\cite{LuPhysicaA11,lerner2005role}.  More recent work has focused on the concept of {\em{role equivalence}}, which relaxes the constraint that equivalent nodes depend on the identical neighbors and requires instead that they depend on other equivalent nodes~\cite{LuPhysicaA11,GlenSIGKDD02,LeichtPRE06,EverettSocNet10}.  Role equivalence gives a more generalized sense of the relationship between nodes by defining equivalence in a self-consistent fashion, but many of these approaches are defined only  for undirected or unweighted networks and do not incorporate externally-defined node attributes (e.g., country or sector information that is available in the WIOD).  In this paper, we develop a measure to identify the most intensive interactions among the GVCs across countries and over time incorporating the full network topology.

To the best of our knowledge, our paper is the first attempt to measure and compare the GVCs at the sector level from a network-based approach. First, from a complex networks perspective, we map the World Input-Output Database (WIOD)~\cite{timmer2015an} into both the upstream and downstream global value networks (GVNs), where the nodes are the individual sectors in different countries and the links are the value-added contribution relationships. Second, we introduce a network-based measure of node similarity to compare the GVCs between any pair of countries for each sector and each year available in the WIOD. Unlike the previous methods, we take into account all the direct and indirect relationships to calculate the GVCs similarity, which provides a more accurate and systemic comparison between the GVCs in space and time.  This measure of similarity may shed light on many important topics of the GVCs, such as dependency, sustainability, risk, and competition associated with the GVCs.  

The rest of the paper is structured as follows. Section 2 describes the WIOD and constructs both the upstream and downstream GVNs and introduces the network-based measure of GVCs similarity. Section 3 summarizes and discusses the results and Section 4 concludes the paper.

\section{Data and Methods}

A network can be broadly defined as a set of items (nodes) and the connections between them (edges)~\cite{albert2002stat,NewmanSIAM03}. 
Recent years have witnessed a burgeoning body of research exploring topics in economics and finance from a network perspective~\cite{pammolli2002,kitsak2010,riccaboni2010,caldarelli2013reconstr,riccaboni2013,zhu2014rise,cerina2015world}. 
The set of sectors and the input-output relationships between them can also be considered as a interdependent network~\cite{cerina2015world}. In this section we first map the WIOD into both the upstream and downstream global value networks (GVNs), where the nodes are the individual sectors in different countries and the links are the value-added contribution relationships. 
Notice that the GVNs are both directed (i.e., links going from value-added provider sectors to receiver sectors) and weighted (i.e., the share of value-added contribution varies from one link to another). 
As a result, the upstream (or downstream) value system of a sector can be obtained by searching for all the direct and indirect incoming (or outgoing) neighbors of the given sector in the upstream (or downstream) GVN. We then propose a measure of GVC similarity that is applicable to the both directed and weighted GVNs with externally defined node attributes (the country and sector of the node) so that we can quantify how similar the GVCs are between any pair of countries for each sector and each year available in the WIOD.

\subsection{Data}

We use the recently available GMRIO database, the WIOD, to investigate the GVCs at the sector level~\cite{tukker2013global}. 
At the time of writing, the WIOD input-output tables cover 35 sectors for each of the 40 economies (27 EU countries and 13 major economies in other regions) plus the rest of the world (RoW) and the years from 1995 to 2011. The 40 economies are representative of the world economy in a sense that they produce around 84.1\% of the world GDP in 2011. 
Table~\ref{Tab:WIOD_ctry} and Table~\ref{Tab:WIOD_ind} list the countries and sectors covered in the WIOD. For each year, there is a harmonized international input-output table listing the input-output relationships between any pair of sectors in any pair of economies. The numbers in the WIOD are in current basic (producers') prices and are expressed in millions of US dollars. 
In a GMRIO table, the input-output flows between sectors is called the transactions matrix and is often denoted by $\boldsymbol{\mathrm{Z}}$. The rows of $\boldsymbol{\mathrm{Z}}$ are the distributions of the sector outputs throughout the two economies, while the columns of $\boldsymbol{\mathrm{Z}}$ are the distributions of inputs required by each sector. Note that sectors often buy inputs from themselves, due to the sector aggregation. Besides intermediate sector use, the remaining outputs are absorbed by the additional columns of final demand, which includes household consumption, government expenditure, etc. 
Similarly, production necessitates not only inter-sectoral transactions but also labor, management, depreciation of capital, and taxes, which are  denoted by the value-added vector $\boldsymbol{\mathrm{v}}$.  The final demand matrix is often denoted by $\boldsymbol{\mathrm{F}}$ and the total sector outputs are denoted by the vector $\boldsymbol{\mathrm{x}}$.

\subsection{Construct the Global Value Networks}

Defining $\boldsymbol{\mathrm{1}}$ a vector of 1's of conformable size (i.e. with the vector length appropriate for the multiplying matrix), and  $\boldsymbol{\mathrm{F}}\cdot\boldsymbol{\mathrm{1}}=\boldsymbol{\mathrm{f}}$, we can write the total global production as the production used for the internal dependencies and the final demand, $\boldsymbol{\mathrm{x}}=\boldsymbol{\mathrm{Z}}\cdot\boldsymbol{\mathrm{1}}+\boldsymbol{\mathrm{f}}$. Dividing each column of $\boldsymbol{\mathrm{Z}}$ by its corresponding total output in $\boldsymbol{\mathrm{x}}$ produces the so-called technical coefficients matrix $\boldsymbol{\mathrm{A}}$, with the terminology signifying that  they represent the technologies employed by the sectors to transform inputs into outputs. Replacing $\boldsymbol{\mathrm{Z}}\cdot\boldsymbol{\mathrm{1}}$ with $\boldsymbol{\mathrm{A}}\boldsymbol{\mathrm{x}}$, we rewrite the output as $\boldsymbol{\mathrm{x}}=\boldsymbol{\mathrm{A}}\boldsymbol{\mathrm{x}}+\boldsymbol{\mathrm{f}}$ and find that $\boldsymbol{\mathrm{x}}={(\boldsymbol{\mathrm{I}}-\boldsymbol{\mathrm{A}})}^{-1}\boldsymbol{\mathrm{f}}$.  
The matrix ${(\boldsymbol{\mathrm{I}}-\boldsymbol{\mathrm{A}})}^{-1}$ is often denoted by $\boldsymbol{\mathrm{L}}$ and is called the Leontief inverse~\cite{leontief1936quantitative,miller2009input}.

Dividing each element of $\boldsymbol{\mathrm{v}}$ by its corresponding total output in $\boldsymbol{\mathrm{x}}$, we define the value-added share vector $\boldsymbol{\mathrm{w}}$.  Defining the operation of a 'hat' over a vector to result in a diagonal matrix with the vector on its diagonal, the value-added contribution matrix can be computed as 
\begin{equation}
	\boldsymbol{\mathrm{G}}=\hat{\boldsymbol{\mathrm{w}}}\boldsymbol{\mathrm{L}}\hat{\boldsymbol{\mathrm{f}}}
\end{equation}
where $\boldsymbol{\mathrm{G}}$ is the value-added contribution matrix and its element $G_{ij}$ is sector $i$'s value-added contribution to sector $j$'s total final demand, $f_j$.  The upstream value-added share matrix, $\boldsymbol{\mathrm{U}}$, is defined as the column-normalized version of $\boldsymbol{\mathrm{G}}$, 
\begin{equation}
	\boldsymbol{\mathrm{U}}=\boldsymbol{\mathrm{G}}{(\widehat{{\boldsymbol{\mathrm{G}}}^T{\boldsymbol{\mathrm{1}}}})}^{-1}
\end{equation}
where the element $U_{ij}$ is sector $i$'s share of value-added contribution out of sector $j$'s total final demand, $f_j$.
The downstream value-added share matrix, $\boldsymbol{\mathrm{D}}$, is similarly defined as the row-normalized version of $\boldsymbol{\mathrm{G}}$:
\begin{equation}
	\boldsymbol{\mathrm{D}}={(\widehat{\boldsymbol{\mathrm{G1}}})}^{-1}\boldsymbol{\mathrm{G}}
\end{equation}
where the element $D_{ij}$ is sector $j$'s share out of sector $j$'s total value-added contribution.  Note that the sum of each column of $\boldsymbol{\mathrm{U}}$ is 1 while the sum of each row of $\boldsymbol{\mathrm{D}}$ is 1.  $\boldsymbol{\mathrm{U}}$ identifies the shares of the value-added providers for any given sector while $\boldsymbol{\mathrm{D}}$ identifies the shares of the value-added receivers for any given sector. Finally, the upstream GVNs are constructed by using $\boldsymbol{\mathrm{U}}$ as the weight matrix while the downstream GVNs can be constructed with $\boldsymbol{\mathrm{D}}$ as the weight matrix. Notice that the GVNs are directed, weighted, and contain self-loops.

\subsection{A Network-Based Measure of Node Similarity}

\begin{figure}[H]
\begin{center}
\includegraphics[width=\textwidth]{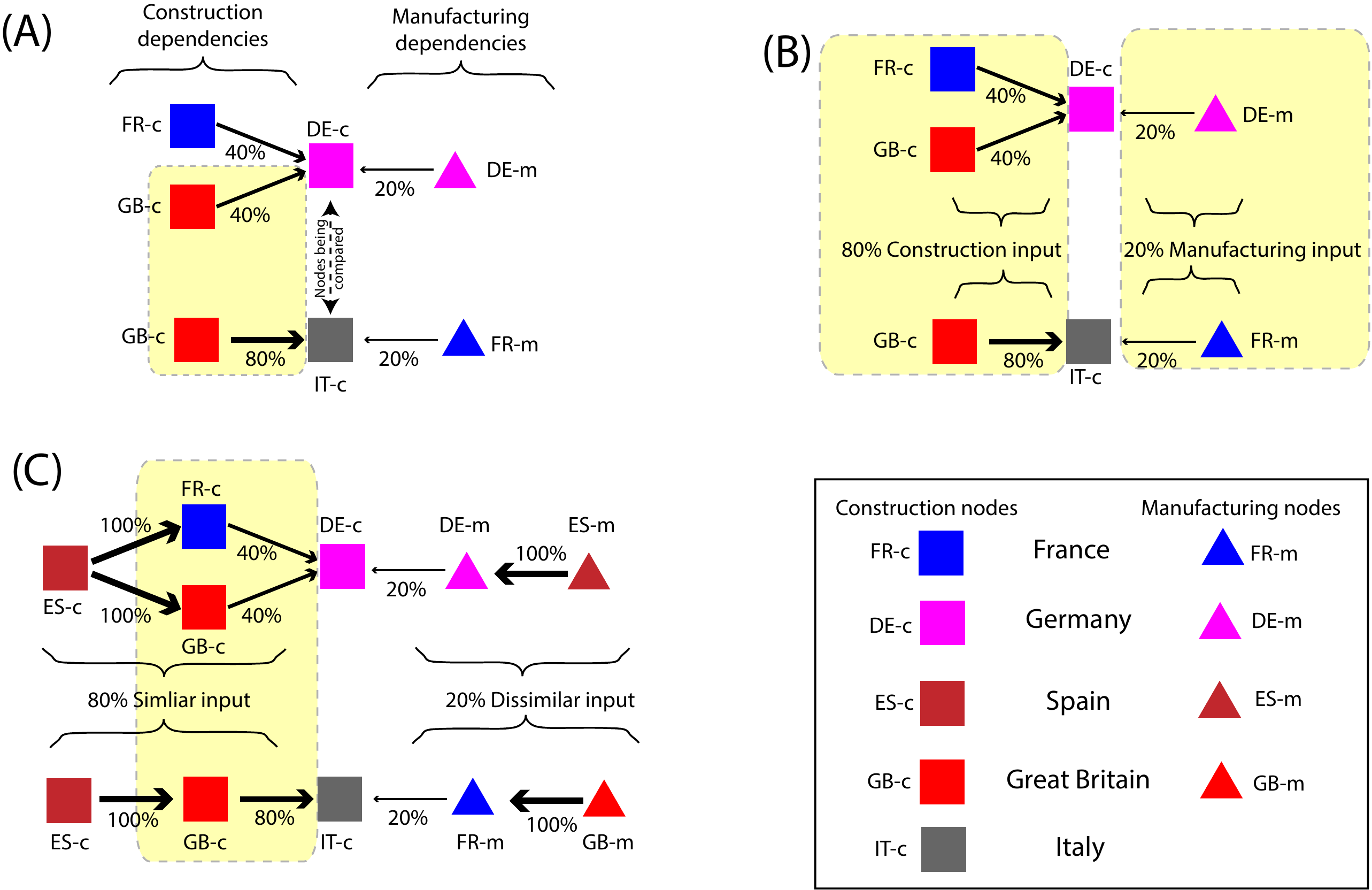}
\caption{Schematic diagrams of the methods of measuring the similarity between two nodes in the GVNs, with hypothetical dependencies of the French and German Construction sectors ($P=$ DE-c and $Q=$ IT-c respectively) shown.  Construction sectors are shown as squares and manufacturing sectors as triangles, while countries are represented by color (France is blue, Germany cyan, Spain brown, Britain red, and Italy gray).  Dependency links that provide a significant contribution to the similarity between DE-c and IT-c are highlighted in yellow.  In (A), we diagram structural similarity using purely local dependency information (as in $J_{PQ}$, $C_{PQ}$, and $S_{PQ}^{(0)}$), with the similarity between DE-c and IT-c due solely to the overlap between the identical provider of British Construction (GB-c).  In (B), we show the sectoral dependency of the nodes are assumed identical (captured in $S_{PQ}^{(1)}$), so all links contribute to the similarity if national differences are ignored.  (C) shows an interpolation between these two extremes, where all upstream construction links for both DE-c and IT-c have the same provider (ES-c), making these providers similar, but the manufacturing links for DE-c and IT-c have different providers. }
\label{MethodSchematic.fig}
\end{center}
\end{figure}

A wide range of similarity measures between nodes in a complex network have been developed recently\cite{LuPhysicaA11} that could potentially be used to determine similar nodes in the GVNs.  The simplest of these that are applicable to weighted networks include those defined by a comparison of the overlap of direct providers, with prominent examples including the weighted Jaccard coefficient\cite{ioffe2010improved} or cosine similarity\cite{LuPhysicaA11} between a pair of nodes $P$ and $Q$ (with each node representing a country-sector pair).  These measures are respectively defined as 

\begin{equation}
J_{PQ}=\sum_{cs}\min(p_{cs},q_{cs})/\sum_{cs}\max(p_{cs},q_{cs})
\end{equation}

and 

\begin{equation}
C_{PQ}=\sum_{cs} p_{cs}q_{cs}/[(\sum_{cs} p_{cs}^2)(\sum_{cs} q_{cs}^2)]^{1/2}
\end{equation}

where $p_{cs}$ is the dependence of $P$ on the country-sector $cs$ (and a similar definition for $q_{cs}$) and the summation runs over all countries $c$ and all sectors $s$ that either $P$ or $Q$ depend on (i.e., all $c$ and $s$ for which $p_{cs}>0$ or $q_{cs}>0$).   
In this paper, we will focus on a different but related similarity measure defined by
\begin{eqnarray}
	S_{PQ}^{(0)}=\frac{\sum_{cs} \left[p_{cs}^2+q_{cs}^2-(p_{cs}-q_{cs})^2\right]}{\sum_{cs} \left[p_{cs}^2+q_{cs}^2+(p_{cs}-q_{cs})^2\right]}\label{firstDef}.
\end{eqnarray}
$J_{PQ}$, $C_{PQ}$ and $S^{(0)}_{PQ}$ share a number of desirable properties in common:  they are all strictly bounded between 0 and 1, with the value 0 attained iff $P$ and $Q$ have no providers in common and the value 1 attained iff $P$ and $Q$ receive from the same nodes by an identical amount.  We further show in the Appendix this definition is strongly related to the definition of the weighted Jaccard Coefficient and differs from the cosine similarity only by a different normalization.  
The general characteristics of these local measures of similarity are schematically diagrammed in Fig. \ref{MethodSchematic.fig} (A) for a hypothetical dependency network of German Construction (node $P$) and Italian Construction (node $Q$).  For all three, only identical dependencies between providing a contribution to the measure of similarity between $P$ and $Q$.    In this hypothetical example, $J_{PQ}=0.25$, $C_{PQ}\approx 0.647$, and $S_{PQ}^{(0)}= 4/9$.  

While purely local measures of similarity have been implemented in a wide range of studies, they are too limited to fully understand the relationship between national production systems because upstream providers that are `similar' but not identical contribute nothing to the measure of similarity between $P$ and $Q$.  
More meaningful information about the similarity between two production systems can be extracted by defining a measure of role equivalence\cite{LuPhysicaA11,GlenSIGKDD02,LeichtPRE06} which implements a more self-consistent measure of similarity.  Existing methods of measuring role equivalence may not be appropriate for the study of the GVCs, because the attributes of each node in the network cannot necessarily be treated on an equal footing.  One might expect that a country-sector pair could change the nationality of its provider (for example, German construction exchanging its direct input from French construction to the construction sector in another nation), but not change the sector of the input (German construction could not replace its French construction input to another industrial sector, regardless of the nation of origin).  The differing economic meanings behind the node attributes suggest that we develop a measure of similarity that explicitly takes these attributes into account (as in $S_{PQ}$).

The definition of $S_{PQ}^{(0)}$ in Fig. \ref{firstDef} represents a lower bound on any meaningful definition of role equivalence between country-sector pairs, because it treats each distinct national production system as {\em{completely}} different.  We can define an upper bound for similarity in a related manner by assuming that national systems of production are all completely identical instead of being completely distinct.  This approximation is schematically diagrammed in Fig. \ref{MethodSchematic.fig} (B), where sectors of production are considered distinct (Construction and Manufacturing are different fields) but national identities are treated as irrelevant.  A measure of similarity equivalent to that in Eq. \ref{firstDef} can be developed in this approximation, with
\begin{eqnarray}
S_{PQ}^{(1)}&=&\frac{\sum_{s} \left[\left(\sum_cp_{cs}\right)^2+\left(\sum_cq_{cs}\right)^2-(\sum_cp_{cs}-q_{cs})^2\right]}{\sum_{s} \left[\left(\sum_cp_{cs}\right)^2+\left(\sum_cq_{cs}\right)^2+(\sum_cp_{cs}-q_{cs})^2\right]}\nonumber\\
&=&\frac{\sum_{cs} \left[p_{cs}^2+q_{cs}^2-(p_{cs}-q_{cs})^2\right]+\sum_sT^-_{PQ}(s)}{\sum_{cs} \left[p_{cs}^2+q_{cs}^2+(p_{cs}-q_{cs})^2\right]+\sum_sT^+_{PQ}(s)}\label{secondDef}
\end{eqnarray}
where we have defined $T^\pm_{PQ}(s)=\sum_{c\ne c'}\left[p_{cs}p_{c's}+q_{cs}q_{c's}\pm (p_{cs}-q_{cs})(p_{c's}-q_{c's})\right]$.  In Fig. \ref{MethodSchematic.fig} (B) it is straightforward to see that $S_{PQ}^{(1)}=1$, because the inputs on the sectoral level are identical between German construction and Italian construction.  We note that Eq. \ref{secondDef} is identical to Eq. \ref{firstDef} in the absence of the terms $T_{PQ}^\pm(s)$ (a fact that is the primary reason for our choice in using this measure of similarity). 

The difference between perfect national similarity (Eq. \ref{secondDef}) and perfect national dissimilarity (Eq. \ref{firstDef}) is entirely contained within the sector-dependent terms $T_{PQ}^{\pm}(s)$, and we note that $T_{PQ}^\pm(s)$ is a sum over terms involving the direct relationship between $P$ and $Q$ to the countries $c$ and $c'$ in sector $s$.  In the context of a role equivalence calculation, these terms should not all be treated equally:  country-sector pairs that are role-equivalent should contribute significantly to the similarity of $P$ and $Q$,  while country-sector pairs that are not role-equivalent should not contribute (diagrammed schematically in Fig. \ref{MethodSchematic.fig} (C)).  This can be accomplished by weighting each term in the sum by the similarity between country $c$ and $c'$ in sector $s$, and we thus write the self-consistent relation
\begin{eqnarray}
	S_{PQ}=\frac{\sum_{s} \sum_{c,c'}\left\{\left[p_{cs}p_{c's}+q_{cs}q_{c's}- (p_{cs}-q_{cs})(p_{c's}-q_{c's})\right]\times S_{cs,c's}\right\}}{\sum_{s} \sum_{c,c'}\left\{\left[p_{cs}p_{c's}+q_{cs}q_{c's}+ (p_{cs}-q_{cs})(p_{c's}-q_{c's})\right]\times S_{cs,c's}\right\}}\label{finalDef}
\end{eqnarray}
as our final expression for the similarity between two country-sectors $P$ and $Q$.  It is straightforward to verify that the diagonal elements identically satisfy $S_{PP}\equiv 1$ for all country sector pairs $P$, and that $S_{PQ}^{(0)}\le S_{PQ}\le S_{PQ}^{(1)}$ for all $P$ and $Q$.  If all countries are treated as different (with $S_{PQ}=0$ for $P\ne Q$) Eq. \ref{finalDef} reduces to Eq. \ref{firstDef}, whereas Eq. \ref{finalDef} reduces to Eq. \ref{secondDef} if all countries are assumed identical (with $S_{PQ}=1$ for all countries).  In the Appendix, we discuss some additional numerical properties of Eq. \ref{finalDef} and the algorithm we use to determine the numerical values of the similarity.  Eq. \ref{finalDef} incorporates a comparison between each of the direct providers of $P$ and $Q$, but by weighting each term by the similarity implicitly includes a comparison between the {\em{indirect}} suppliers of $P$ and $Q$ (those that are providers of the providers).  Two different direct providers of $P$ and $Q$ that themselves have similar inputs will have a large contribution to the similarity $S_{PQ}$, while direct providers who themselves have very different value chains will give a small contribution.  
This can be clearly seen by computing the similarity in Fig. \ref{MethodSchematic.fig} (C), where we numerically find $S_{PQ}\approx 0.889$ (in comparison to $S_{PQ}^{(0)}\approx 0.444$ and $S_{PQ}^{(1)}=1$).  
This shows that Eq. \ref{finalDef} captures our expectation that the similarities in the direct construction inputs due to the shared indirect link (Spanish construction) increases the similarity between German and Italian construction, but the dissimilarities in the direct manufacturing suppliers prevent a perfect role-similarity between them. 

The magnitude of $S_{PQ}$ by itself cannot distinguish between similarity due to $P$ and $Q$ sharing identical providers versus sharing role-equivalent providers, we further define the rescaled similarity 
\begin{eqnarray}
	R_{PQ}=\frac{S_{PQ}-S^{(0)}_{PQ}}{S^{(1)}_{PQ}-S^{(0)}_{PQ}}
	\label{rescaleDef}
\end{eqnarray}
which indicates how close $S_{PQ}$ is to its upper bound $S^{(1)}_{PQ}$. Because the upper bound $S^{(1)}_{PQ}$ completely ignores the national difference, if $R_{PQ}$ is very close to 1, it means that there is a significant national similarity between the sectors compared. In other words, the rescaled version allows us to attribute its magnitude to the national similarity of different nations.

In this section we have only discussed the similarity based on the upstream GVNs, whose adjacency matrix $\boldsymbol{\mathrm{U}}$ is both asymmetrical (directed) and real-valued between 0 and 1 (weighted) and with non-zero diagonal elements (self-loops). Measuring a downstream similarity using the methods in this section can be equivalently accomplished by applying the same methodologies to the transposed downstream networks (reversing the direction of the links, so that receiver sectors become provider sectors).

\section{Results}

\subsection{General Patterns of Similarity}

\begin{figure}[H]
 \begin{center}
  \includegraphics[width=\textwidth]{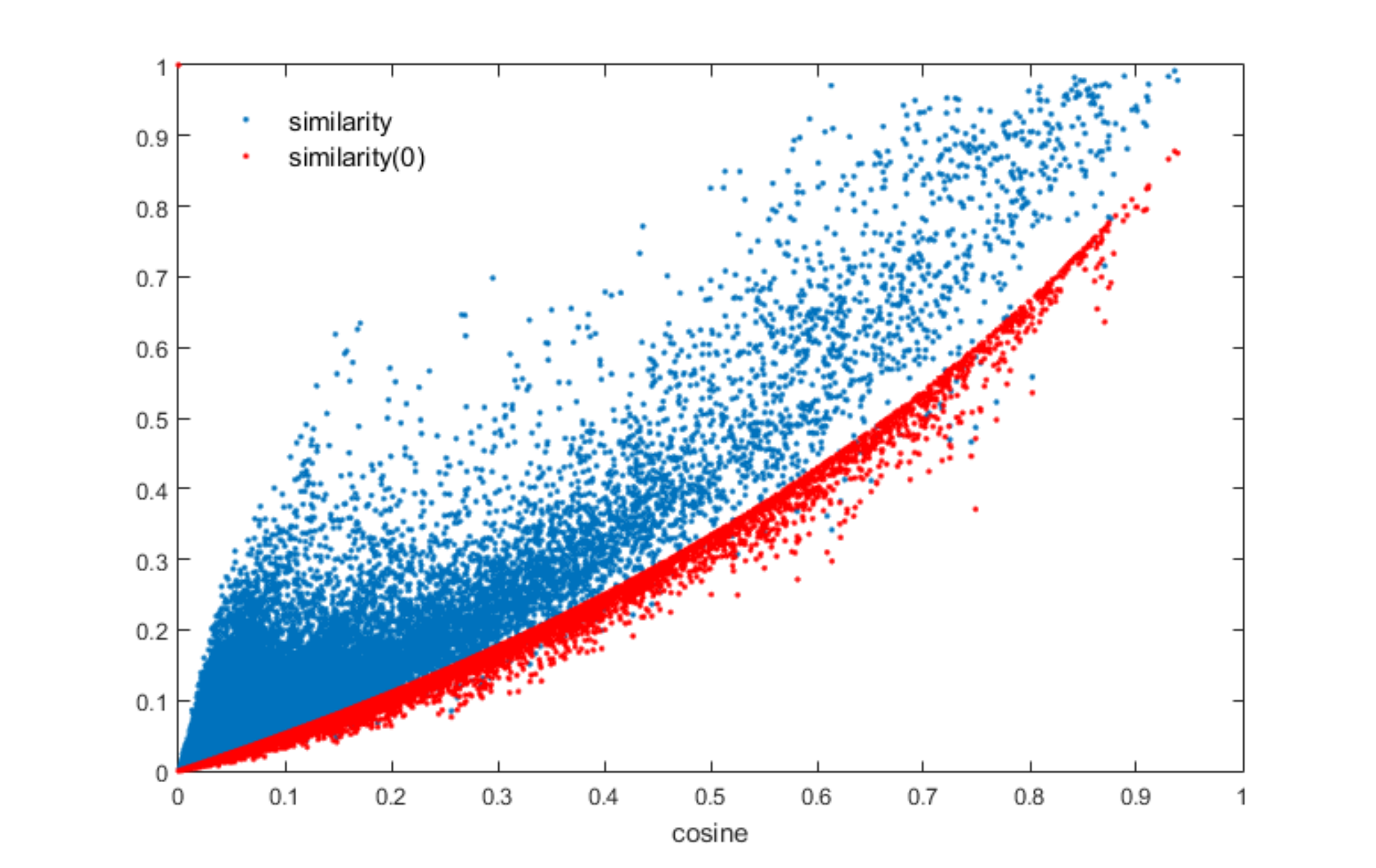}
  \caption{The scatter plot of cosine similarity (x-axis) vs. our measures of similarity for all pairwise
	comparison of sectors across countries for the upstream GVNs and for all years. Both $S^{(0)}_{PQ}$ and $S_{PQ}$ are reported in red and blue respectively.}
  \label{overlap}
 \end{center}
\end{figure}

We compute the pairwise similarity across countries for each industry and each year available in the WIOD. It is worthwhile to examine how strongly correlated our measure of similarity is with other alternative measures. They tend to be highly correlated, with the rescaled version of our measure of similarity more highly correlated with $C_{PQ}$ ($.83$ upstream, $.78$ downstream) than with $J_{PQ}$ ($.66$ upstream, $.72$ downstream).  Even though the correlation is high, it must be noticed that, unlike other local measures of similarity, our measure of similarity takes into account both direct and indirect relationships along the value chain.  In Fig. \ref{overlap}, we see that when we include indirect value-added providers in the computation of the upstream similarity, country-sector pairs become more similar to one another. Our network-based measure of similarity is much less correlated with cosine similarity than the local version (the lower bound $S_{PQ}^{(0)}$) of our index, which differs from cosine only in the normalization term.  

\begin{figure}[H]
 \begin{center}
  \includegraphics[width=\textwidth]{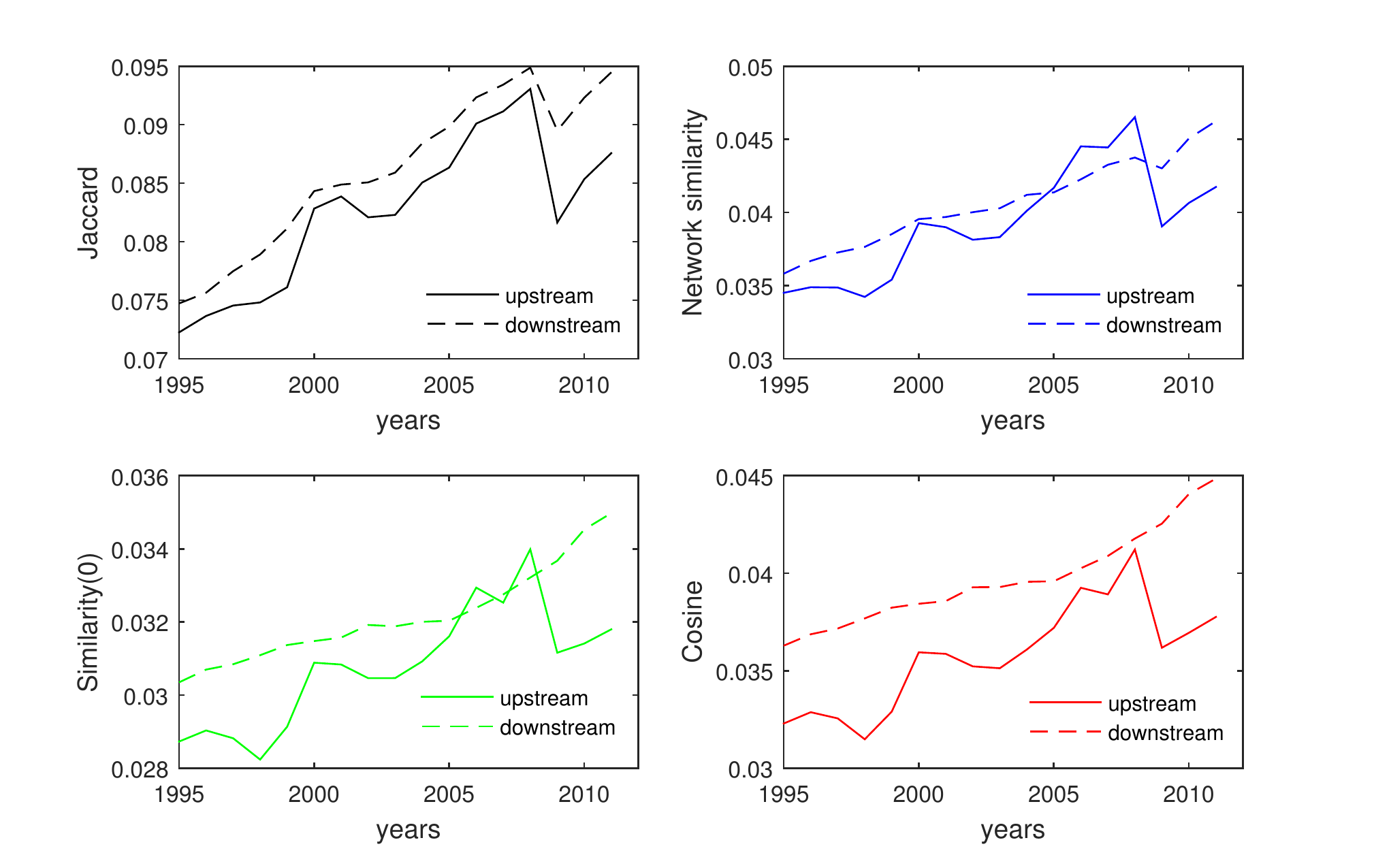}
  \caption{The evolution of average similarity across countries and sectors over time, 1995-2011. 
	  We compare four different measures of similarity: Jaccard [$J_{PQ}$], Cosine [$C_{PQ}$], Similarity(0) [$S^{(0)}_{PQ}$], and Network Similarity [$S_{PQ}$]. 
	For every indicator, we report both upstream (solid lines) and downstream (broken lines) similarity.}
  \label{trend}
 \end{center}
\end{figure}

We explore the evolution of the similarity between sectors by computing the mean similarity for all sectors and country pairs, $\sum_s\sum_{c\ne c'}S_{cs,c's}/N_sN_c(N_c-1)$, with $N_c=41$ the number of countries and $N_s=35$ the number of sectors.   Fig. \ref{trend} reveals that, on average sectors across the globe tend to be more similar over time, a fact that is consistently observed using all measures of similarity.  All measures also show that upstream similarity is more volatile and less intense than the downstream similarity. However, when all network interdependences are
taken into account, the upstream and downstream similarities tend to more closely follow the same path of growth and both exhibit a temporary reduction in the aftermath of the great recession in 2008 (the latter is also captured by Jaccard).

\begin{figure}[H]
 \begin{center}
  \includegraphics[width=\textwidth]{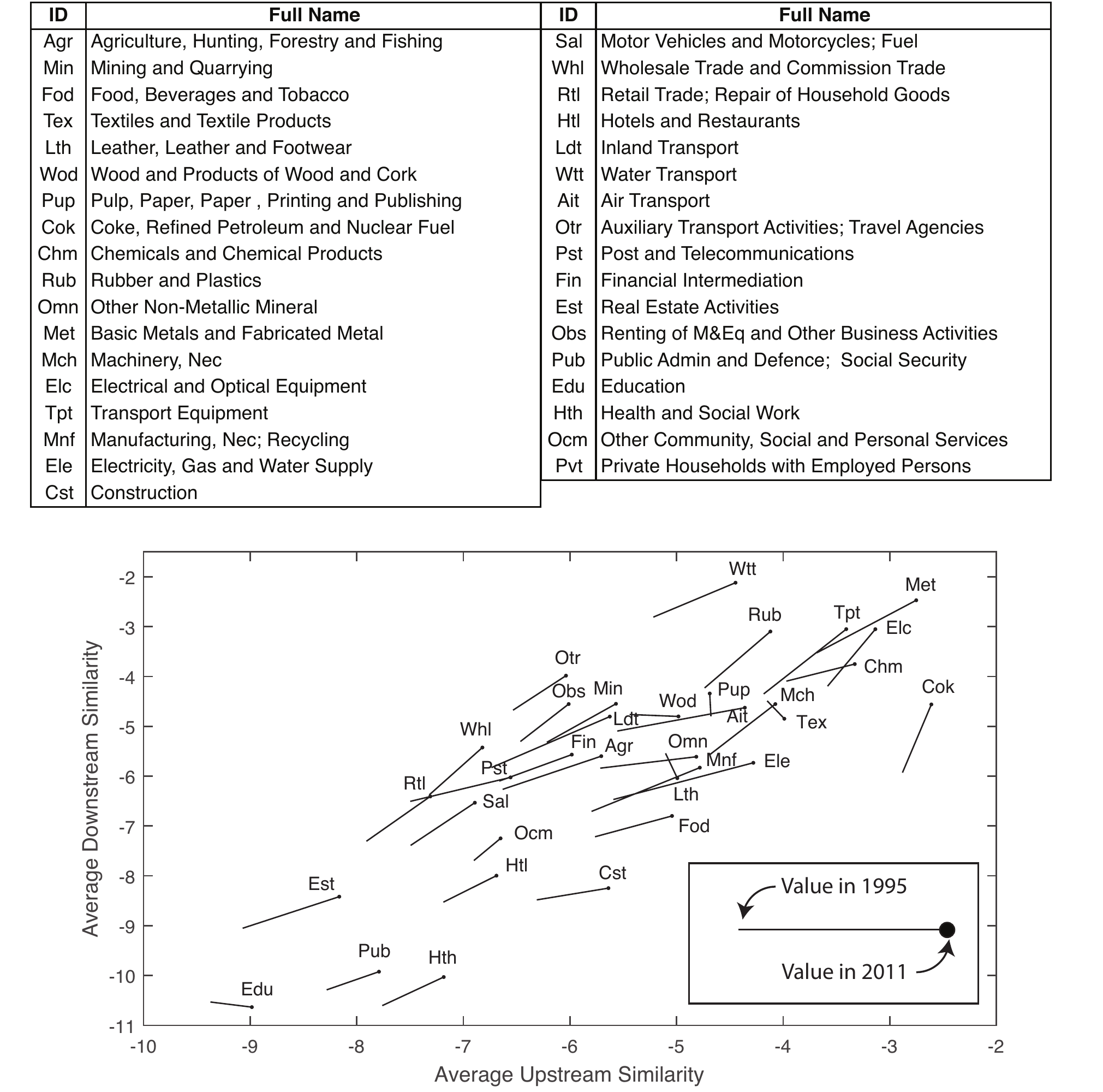}
  \caption{The average upstream and downstream similarities of sectors for the years 1995 and 2011 using a logarithmic scale.}
  \label{sectorsim}
 \end{center}
\end{figure}

For each year, we can average across countries to have the average similarity for each industry, $\sum_{c\ne c'}S_{cs,c's}/N_c(N_c-1)$. Fig. \ref{sectorsim} shows both the average upstream and downstream similarities for all the sectors and for the years 1995 and 2011. It is straightforward to see that most sectors have increased their similarities over time as most ``arrows'' are pointing to the northeast direction. 
Sectors like ``Coke, Refined Petroleum and Nuclear Fuel (Cok)'' have high average upstream similarity and relatively low average downstream similarity, which means that it is more likely to find country-sector overlap in their upstream value chains. This makes sense for the sector ``Cok'': energy providers tend to be concentrated in only a few countries. More generally, the manufacturing sectors tend to be more similar across countries than the services sectors as the former is clustered in the top right of Fig. \ref{sectorsim} and the latter is clustered in the lower left of Fig. \ref{sectorsim}.

\begin{figure}[H]
 \begin{center}
  \includegraphics[width=\textwidth]{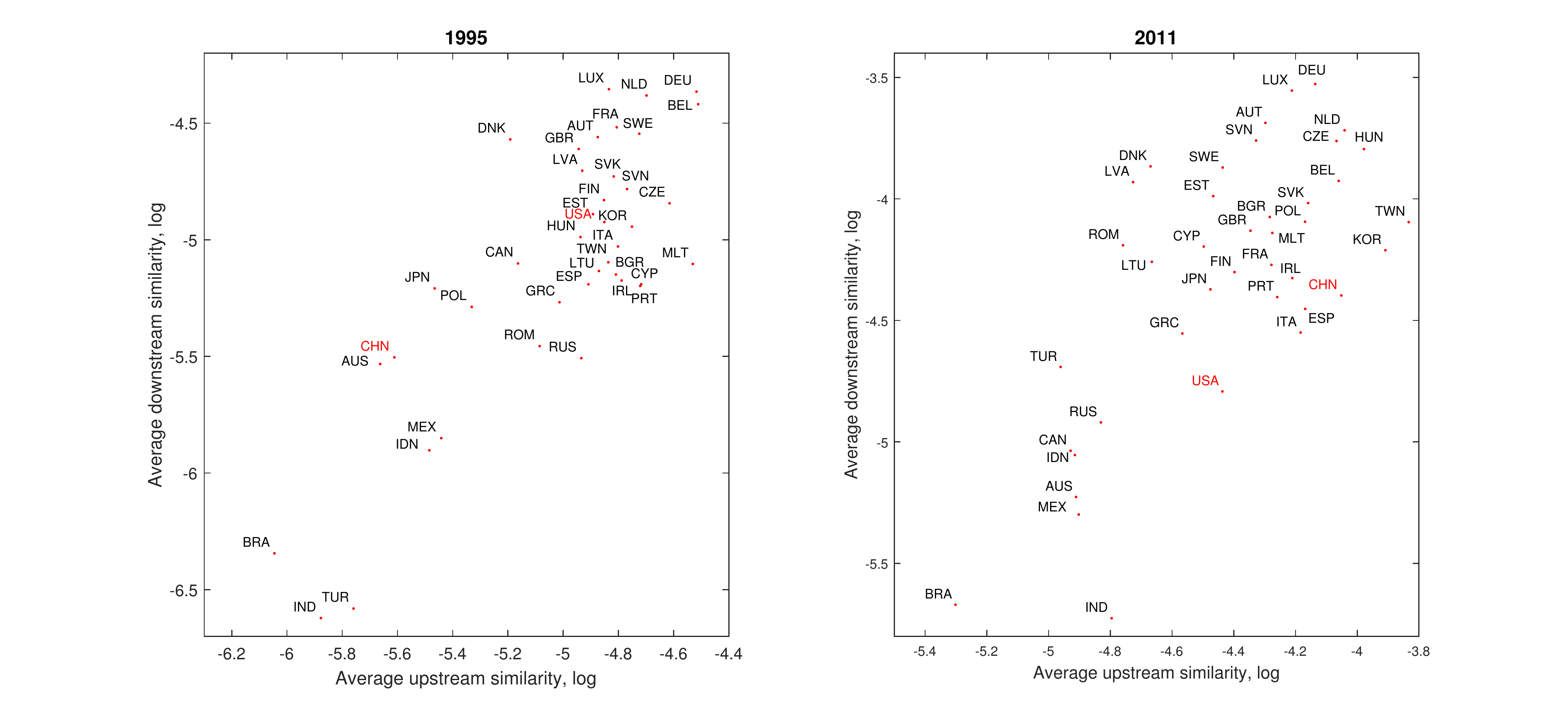}
  \caption{The average upstream and downstream similarities of countries in years 1995 and 2011 and using logarithmic axes.}
  \label{countrysim}
 \end{center}
\end{figure}

For each year, we can also average across industries and foreign economies ($\sum_s\sum_{c\ne c'}S_{cs,c's}/N_s(N_c-1)$) to define an mean similarity for each nation. Fig.~\ref{countrysim} shows both the average upstream and downstream similarities for all the countries and for the years 1995 and 2011. Again, we observe a general increasing trend of the similarities (see the change of the axis range over time).
Furthermore, the ``Asian miracle'' economies, South Korea and Taiwan, are clearly associated with high average similarities when compared with other countries. 
As in the study of Ref. \cite{dean2011measuring}, we also find that China has been increasingly involved in the vertical specialization and has made a dramatic move over time that it has joined the other ``Asian miracle'' nations in terms of the similarities. In the Appendix, we further report the clustering results based on the average similarities of countries.

\subsection{Specific Case Studies}

\begin{figure}
 \begin{center}
  \includegraphics[width=.9 \textwidth]{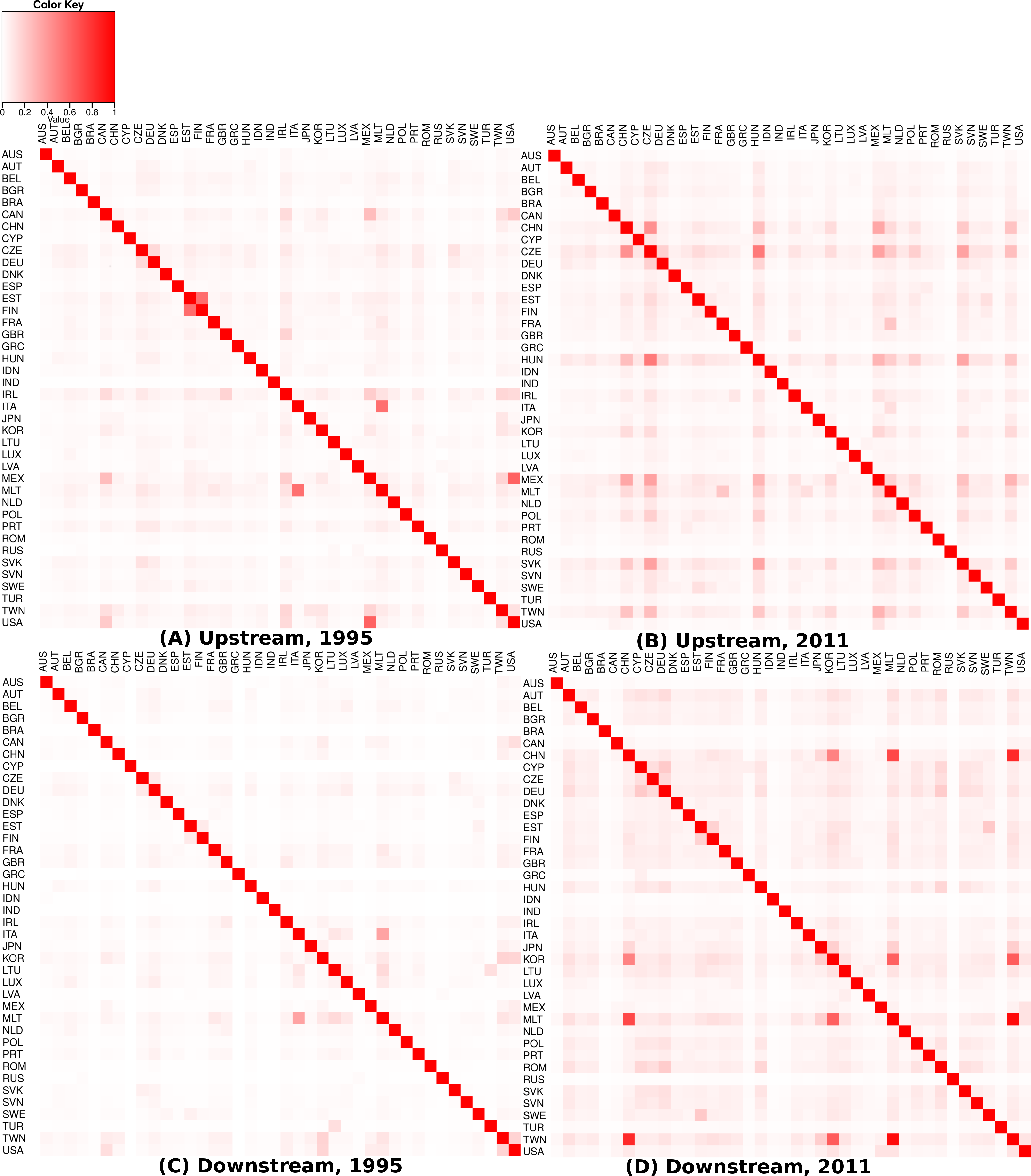}
  \caption{The pairwise upstream and downstream similarity across countries of the electrical equipment sector in 1995 and 2011. 
  Darker color indicates higher values. In 1995, China is not very similar to any other countries. In 2011, the most similar countries to China are Czech Republic (upstream) and Taiwan (downstream).} 
  \label{fig:upstream}
 \end{center}
\end{figure}

A convenient way to organize our results is to show the country-by-country matrix of pairwise similarities for specific sectors and years. Fig.~\ref{fig:upstream} is an example for the upstream rescaled similarity and the downstream rescaled similarity for the electrical engineering sector, ``Elc'' (see~\cite{de2013mapping,ferrarini2011} for a recent analysis of the same sector).  Notice that, by our definition of similarity, the matrix is symmetrical and has all 1's in its diagonal.  There is a visually clear increase in the similarity between most nations in ``Elc'' between 1995 and 2011, and many economies that were very dissimilar in 1995 became very similar in 2011 (with China being a prominent example).  In 1995, China is neither upstream-similar nor downstream-similar to any other countries as its corresponding rows or columns are barely colored. In 2011, however, China becomes fairly upstream-similar to Czech Republic, Hungary, Mexico, Slovak, Taiwan, etc, with Czech Republic as its most upstream-similar country. On the other hand, China becomes highly downstream-similar to South Korea and Taiwan, with Taiwan as its most downstream-similar country.   

\begin{figure}
 \begin{center}
  \includegraphics[width=\textwidth]{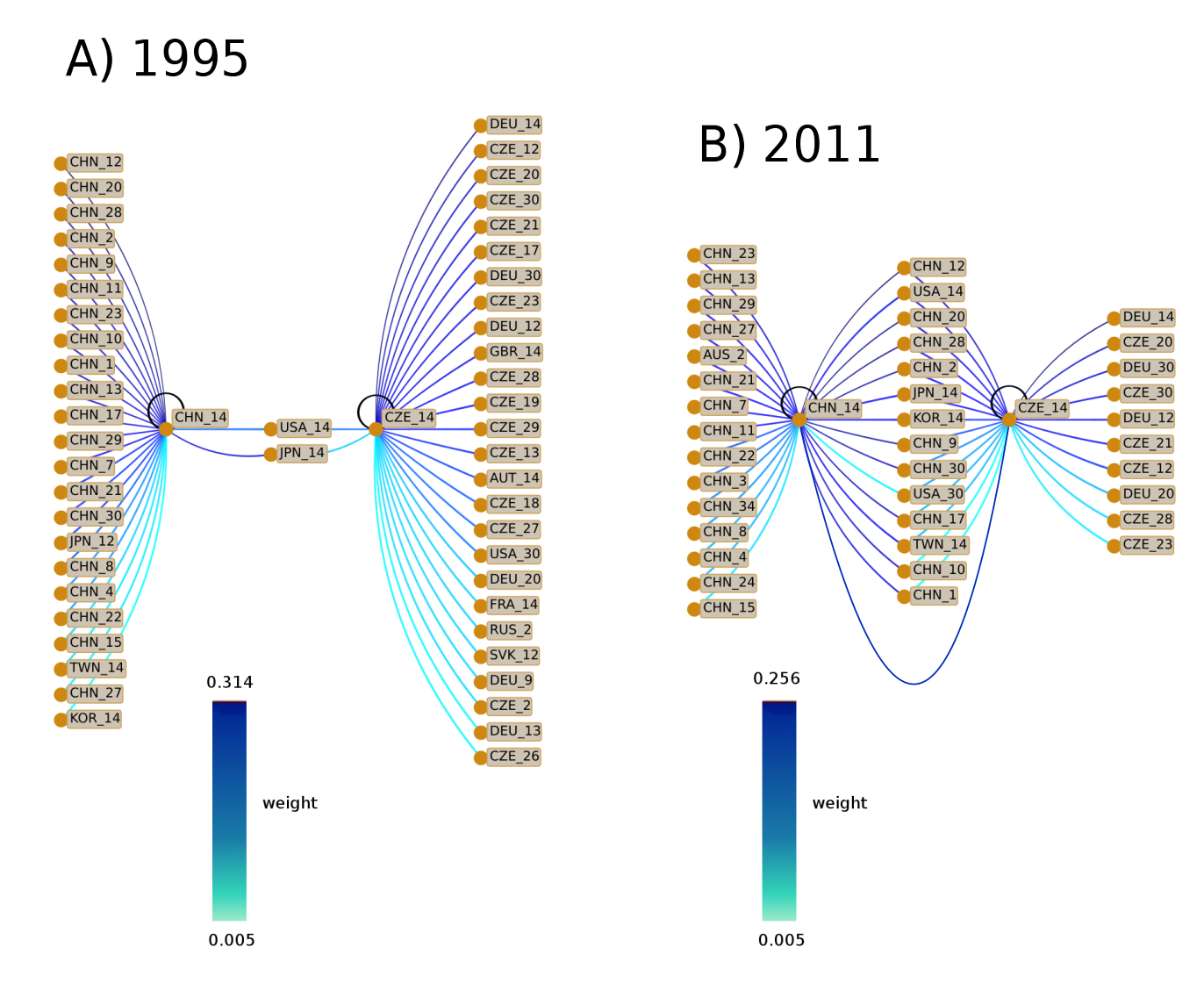}
  \caption{The first-degree neighbors of the electrical equipment sector in China and Czech Republic in the upstream GVNs in 1995 and 2011. Any incoming links to the two sectors with weight greater than or equal to 0.005 are shown. Over time, the number of shared value-added providers increases for the two sectors, and the direct interaction between the two sectors becomes significant as a new link is formed between them in 2011. } 
  \label{fig:CZE_CHN}
 \end{center}
\end{figure}

\begin{figure}
 \begin{center}
  \includegraphics[width=\textwidth]{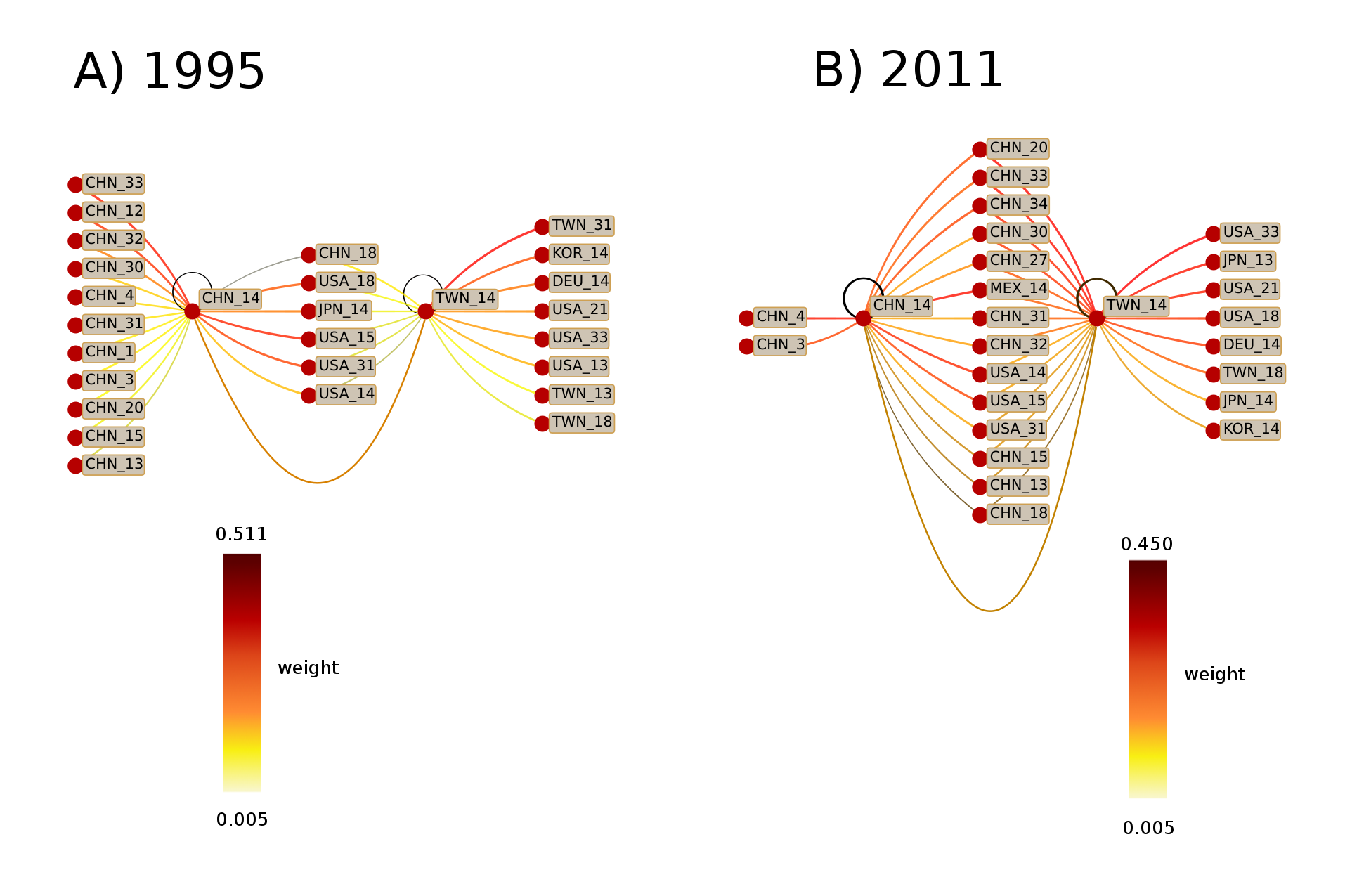}
  \caption{The first-degree neighbors of the electrical equipment sector in China and Taiwan in the downstream GVNs in 1995 and 2011. Any outgoing links from the two sectors with weight greater than or equal to 0.005 are shown. Over time, the number of shared value-added receivers has increased for the two sectors.} 
  \label{fig:TWN_CHN}
 \end{center}
\end{figure}

To see the dynamics at a finer resolution, we show the significant first-degree neighbors (i.e., those with link weight no less than 0.005) of the electrical equipment sector in China and Czech Republic in the upstream GVNs in 1995 and 2011 in Fig. \ref{fig:CZE_CHN}, and in Fig.~\ref{fig:TWN_CHN} the significant first-degree neighbors of ``Elc'' in China and Taiwan in the downstream GVNs in 1995 and 2011. Note that while our measure of similarity takes into account all the indirect neighbors, we only show the first-degree neighbors in Figs. \ref{fig:CZE_CHN}-\ref{fig:TWN_CHN} for better visualization.  Over time, the number of shared value-added providers between China and the Czech Republic has increased, and a direct interaction between the two sectors becomes significant as a new link is formed between them.  Likewise, the number of shared value-added receivers increases between China and Taiwan over time.

\begin{figure}[H]
 \begin{center}
  \includegraphics[width=\textwidth]{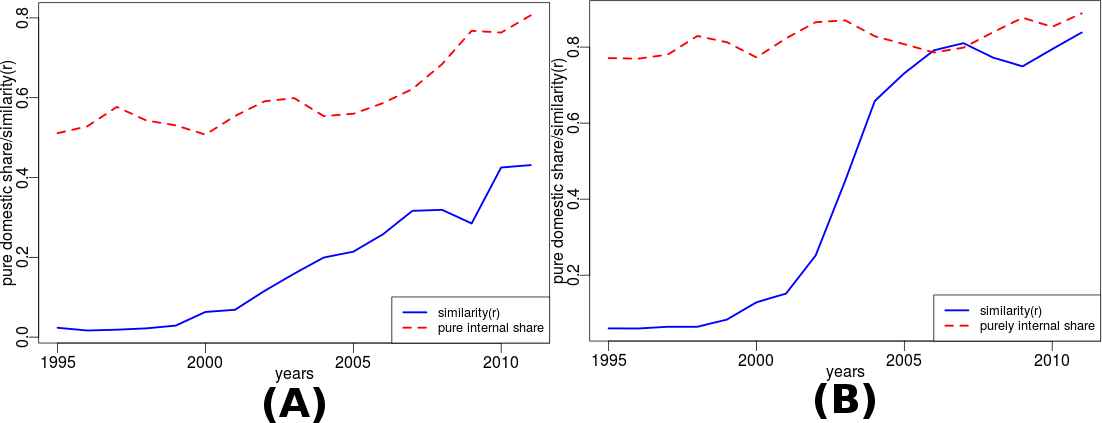}
  \caption{(A) The purely internal share of the upstream similarity (dashed line) and the rescaled upstream similarity (solid line) between the electrical equipment sector (``Elc'') in China and the Czech Republic from 1995-2011. (B) The purely internal share of the downstream similarity and the rescaled downstream similarity between the electrical equipment sector in China and Taiwan from 1995-2011. The purely internal share is well correlated with the rescaled upstream similarity between China and the Czech Republic, but they are not correlated in the downstream case between China and Taiwan.}  
  \label{fig:domContr}
 \end{center}
\end{figure}

Eq.~\ref{finalDef} can give valuable insights into which sectors are responsible for the increased similarity. We can decompose the numerator of Eq. \ref{finalDef} into individual terms and examine exactly how much each pair of country-sectors contributes to the similarity score. We divide the country-pairs into three categories:  purely internal (both countries either China or Czech Republic for the upstream case), purely external (neither country China nor Czech Republic for the upstream case), and mixed (one either China or Czech Republic and the other a different country for the upstream case).  Fig.~\ref{fig:domContr} (A) shows the purely internal share of the upstream similarity (dashed lines) and the rescaled upstream similarity between the electrical equipment sector in China and the one in Czech Republic over time. 
The purely internal share is well correlated with the upstream similarity in this case (both are increasing in time), which implies that more intensive direct interaction between China and the Czech republic is the main driving force behind their increased similarity. This is indeed supported by Fig.~\ref{fig:CZE_CHN} (A), where the electrical equipment sectors in China and the Czech Republic form a significant direct link between themselves in 2011.  Fig.~\ref{fig:domContr} (B) shows the purely internal share of the downstream similarity and the rescaled downstream similarity between the electrical equipment sector in China and the one in Taiwan over time.  
Unlike the upstream case between China and Czech Republic, the purely internal share is not well correlated with the rescaled downstream similarity, suggesting that the overlap of foreign sectors is likely responsible for their increased similarity instead of shared internal connections.

\section{Concluding remarks}

In recent decades, international trade has been marked by the spatial fragmentation of production, which is captured by the notion of global value chains (GVCs). 
A good understanding of the evolution of the GVCs is of vital importance for the macro decision makers to design proper and timely policies and for the micro decision makers to engage in and benefit from the revolution~\cite{johnson2014}. A method of measuring and comparing the GVCs in a systematic way is necessary for informed decisions on both scales, but about which the existing literature remains silent.  This paper has aimed to fill this gap in the literature. First, we use the World Input-Output Database (WIOD) to construct both the upstream and downstream global value networks where the nodes are the individual sectors in different countries and the links are the value-added contribution relationships. Second, to systematically compare the GVCs, we define a network-based measure of role equivalence that takes the differing types of attributes of each node into account.  Our measure of similarity assumes that while it is possible to exchange the nationality of a direct provider in a particular sector, the sectors themselves are not interchangeable.  Coupling this expectation with naturally-defined lower and upper bounds on similarity permitted the self-consistent definition of similarity.


We have found that manufacturing sectors tend to be more similar across countries than the services sectors while countries like China has increased its average similarity over time. As a case study, we found that the sector of electrical equipment in China has become upstream-similar to the one in Czech Republic and downstream-similar to the one in Taiwan.  Our measure of similarity enables us to identify the most intensive interactions among the GVCs across countries and over time. However, the driving forces behind the interactions can be either internal or external, which can be interpreted as value-chain integration or value-chain competition accordingly.  Identifying and quantifying these differences will be left for future work.

Regarding the potential uses and policy implications of our measure of GVCs similarity, we expect that the GVC similarity will be a better measure than the export/import similarity (measured without reference to the topology of the global network). The latter has been largely used in the trade literature as a proxy for competition and trade diversion between countries. However, the gross trade statistics can be seriously flawed (by double counting) as the global production sharing has become a norm. In addition, the trade diversification measured by the export/import similarity has become a less reliable indicator of a country's competitiveness because similar GVCs are compatible with very dissimilar export outputs (as was the case for China).  Our measure may also be useful as a predictor for future link formation using the link prediction literature in the field of complex networks~\cite{LuPhysicaA11,LuPRE09,ZhouEPJB09}, where high similarity between the country-sector pairs identified by our measure may suggest an increasingly intense value-added relationships in the future.  
Finally, since the GVCs tend to become more similar over time and countries tend to become more vertically specialized, there are concerns about the systemic risk of the global production system. Integration and diversification are two important features for the stability of input-output systems~\cite{elliott2014financial}.  Our results suggest that effective diversification is lower than expected due to the increasing overlap of trading partners along value chains, and hence increases the risk of instability. 

Some possible future extensions to this paper include quantifying the driving forces behind the dynamics of similarity, as mentioned in the previous section.  Our approach can be generalized to networks with more than two types of node attributes, and so long as it is possible to meaningfully define the lower and upper bounds on the similarity given the constraints of the differing attributes it may be of interest to define a similar measure of self-consistent similarity.  This approach can also be modified to incorporate other economically relevant information.  For example, the greater reliance that a sector typically has on itself and the domestic economy at large (in comparison to foreign sectors) may suggest that differentiating between domestic and foreign sectors and treating self-loops differently may be appropriate.  In these cases, adapting the upper and lower bounds found in Eq. \ref{secondDef} and \ref{firstDef} to meaningfully capture the differences between foreign and domestic or between self- and non-self-dependence should naturally give rise to an equivalent self-consistent measure of similarity.

\section{Appendix}

\setcounter{table}{0}
\renewcommand\thetable{A\arabic{table}}

\setcounter{figure}{0}
\renewcommand\thefigure{A\arabic{figure}}

\subsection{WIOD Coverage}

\begin{table}[H] 
  \centering
  \caption{{\bf List of WIOD economies.}}
  \resizebox{\textwidth}{!}{
    \begin{tabular}{@{\extracolsep{4pt}}llllllllll@{}}
    \toprule
    \multicolumn{2}{c}{\textbf{Euro-Zone}} & \multicolumn{2}{c}{\textbf{Non-Euro EU}} & \multicolumn{2}{c}{\textbf{NAFTA}} & \multicolumn{2}{c}{\textbf{East Asia}} & \multicolumn{2}{c}{\textbf{BRIIAT}} \\
    \cline{1-2} \cline{3-4} \cline{5-6}  \cline{7-8} \cline{9-10}
    \textbf{Economy} & \textbf{3L Code} & \textbf{Economy} & \textbf{3L Code} & \textbf{Economy} & \textbf{3L Code} & \textbf{Economy} & \textbf{3L Code} & \textbf{Economy} & \textbf{3L Code} \\
    
    Austria & AUT   & Bulgaria & BGR   & Canada & CAN   & China & CHN   & Australia & AUS \\
    Belgium & BEL   & Czech Rep. & CZE   & Mexico & MEX   & Japan & JPN   & Brazil & BRA \\
    Cyprus & CYP   & Denmark & DNK   & USA   & USA   & South Korea & KOR   & India & IND \\
    Estonia & EST   & Hungary & HUN   &       &       & Taiwan & TWN   & Indonesia & IDN \\
    Finland & FIN   & Latvia & LVA   &       &       &       &       & Russia & RUS \\
    France & FRA   & Lithuania & LTU   &       &       &       &       & Turkey & TUR \\
    Germany & DEU   & Poland & POL   &       &       &       &       &       &  \\
    Greece & GRC   & Romania & ROM   &       &       &       &       &       &  \\
    Ireland & IRL   & Sweden & SWE   &       &       &       &       &       &  \\
    Italy & ITA   & UK    & GBR   &       &       &       &       &       &  \\
    Luxembourg & LUX   &       &       &       &       &       &       &       &  \\
    Malta & MLT   &       &       &       &       &       &       &       &  \\
    Netherlands & NLD   &       &       &       &       &       &       &       &  \\
    Portugal & PRT   &       &       &       &       &       &       &       &  \\
    Slovakia & SVK   &       &       &       &       &       &       &       &  \\
    Slovenia & SVN   &       &       &       &       &       &       &       &  \\
    Spain & ESP   &       &       &       &       &       &       &       &  \\
    \bottomrule
    \end{tabular}%
    }
  \label{Tab:WIOD_ctry}%
\end{table}%

\begin{table}[H]
  \centering
  \caption{{\bf List of WIOD sectors.}}
  \resizebox{\textwidth}{!}{
    \begin{tabular}{@{\extracolsep{4pt}}llll@{}}
    \toprule
    \textbf{Full Name} & \textbf{ISIC Rev. 3 Code} & \textbf{WIOD Code} & \textbf{3-Letter Code} \\
    \cline{1-1} \cline{2-2} \cline{3-3} \cline{4-4}
    Agriculture, Hunting, Forestry and Fishing & AtB   & c1    & Agr \\
    Mining and Quarrying & C     & c2    & Min \\
    Food, Beverages and Tobacco & 15t16 & c3    & Fod \\
    Textiles and Textile Products & 17t18 & c4    & Tex \\
    Leather, Leather and Footwear & 19    & c5    & Lth \\
    Wood and Products of Wood and Cork & 20    & c6    & Wod \\
    Pulp, Paper, Paper , Printing and Publishing & 21t22 & c7    & Pup \\
    Coke, Refined Petroleum and Nuclear Fuel & 23    & c8    & Cok \\
    Chemicals and Chemical Products & 24    & c9    & Chm \\
    Rubber and Plastics & 25    & c10   & Rub \\
    Other Non-Metallic Mineral & 26    & c11   & Omn \\
    Basic Metals and Fabricated Metal & 27t28 & c12   & Met \\
    Machinery, Nec & 29    & c13   & Mch \\
    Electrical and Optical Equipment & 30t33 & c14   & Elc \\
    Transport Equipment & 34t35 & c15   & Tpt \\
    Manufacturing, Nec; Recycling & 36t37 & c16   & Mnf \\
    Electricity, Gas and Water Supply & E     & c17   & Ele \\
    Construction & F     & c18   & Cst \\
    Sale, Maintenance and Repair of Motor Vehicles and Motorcycles; Retail Sale of Fuel & 50    & c19   & Sal \\
    Wholesale Trade and Commission Trade, Except of Motor Vehicles and Motorcycles & 51    & c20   & Whl \\
    Retail Trade, Except of Motor Vehicles and Motorcycles; Repair of Household Goods & 52    & c21   & Rtl \\
    Hotels and Restaurants & H     & c22   & Htl \\
    Inland Transport & 60    & c23   & Ldt \\
    Water Transport & 61    & c24   & Wtt \\
    Air Transport & 62    & c25   & Ait \\
    Other Supporting and Auxiliary Transport Activities; Activities of Travel Agencies & 63    & c26   & Otr \\
    Post and Telecommunications & 64    & c27   & Pst \\
    Financial Intermediation & J     & c28   & Fin \\
    Real Estate Activities & 70    & c29   & Est \\
    Renting of M\&Eq and Other Business Activities & 71t74 & c30   & Obs \\
    Public Admin and Defence; Compulsory Social Security & L     & c31   & Pub \\
    Education & M     & c32   & Edu \\
    Health and Social Work & N     & c33   & Hth \\
    Other Community, Social and Personal Services & O     & c34   & Ocm \\
    Private Households with Employed Persons & P     & c35   & Pvt \\
    \bottomrule
    \end{tabular}%
    }
  \label{Tab:WIOD_ind}%
\end{table}%

\subsection{Relationship with Jaccard and Cosine Similarities}

There are many possible ways of measuring the similarity between nodes in a weighted network using information involving only their nearest neighbors, with the Jaccard\cite{ioffe2010improved} and Cosine\cite{LuPhysicaA11} similarities being often used.  We have chosen to use Eq. \ref{firstDef}, and in this section we show its relationship to both the Jaccard and Cosine similarities.  It is a mathematical identity that
\begin{eqnarray}
J_{PQ}=\frac{\sum_{cs}\min(p_{cs},q_{cs})}{\sum_{cs}\max(p_{cs},q_{cs})}=\frac{\sum_{cs}\left[ p_{cs}+q_{cs}-|p_{cs}-q_{cs}|\right]}{\sum_{cs}\left[ p_{cs}+q_{cs}+|p_{cs}-q_{cs}|\right]}
\end{eqnarray}
with the numerator and denominator differing only in a change of sign on the terms involving the absolute value of $p_{cs}-q_{cs}$.  $J_{PQ}$ satisfies the useful property that $0\le J_{PQ}\le 1$ with the equalities occurring iff $P$ and $Q$ have either no weight to identical nodes or all identical weights.  Many other functional forms satisfy this requirement, though, with a family of examples being 
\begin{eqnarray}
J_{PQ}^{(\alpha)}=\frac{\sum_{cs}\left[ p_{cs}^\alpha+q_{cs}^\alpha-|p_{cs}-q_{cs}|^\alpha\right]}{\sum_{cs}\left[ p_{cs}^\alpha+q_{cs}^\alpha+|p_{cs}-q_{cs}|^\alpha\right]}\label{JaccardP}
\end{eqnarray}
for all $\alpha>0$, with Eq. \ref{firstDef} coinciding with the choice of $\alpha=2$. Due to the convenient link between Eqs. \ref{firstDef} and \ref{secondDef} that could exist only with the choice of $\alpha=2$, there is utility in selecting this specific value of $\alpha$.  
We further note that for $\alpha=2$, the numerator of Eq. \ref{JaccardP} is $\sum_{cp} p_{cp}^2+q_{cp}^2-(p_{cp}-q_{cp})^2=2\sum_{cp} p_{cp}q_{cp}$, exactly twice  the numerator in the definition of Cosine similarity.  
While $S^{(0)}_{PQ}$ and $C_{PQ}$ have differing normalizations, we naturally expect that these measures of similarity will be highly correlated.  The high degree of similarity between the definitions of $S_{PQ}^{(0)}$, $J_{PQ}$, and $C_{PQ}$ suggests that the usage of $S_{PQ}^{(0)}$ is reasonable as a measure of similarity.

\subsection{Computational Algorithm}

The definition of similarity in Eq. \ref{finalDef} is not analytically tractable due to its nonlinearity, and approximate methods for determining the similarity between countries in specific sectors.  We use an iterative method to solve for $S_{PQ}$, by defining the $(k+1)^{th}$ iteration of the similarity as 
\begin{eqnarray}
	S_{PQ;k+1}=\frac{\sum_{s} \sum_{c,c'}\left\{\left[p_{cs}p_{c's}+q_{cs}q_{c's}- (p_{cs}-q_{cs})(p_{c's}-q_{c's})\right] S_{cs,c's;k}\right\}}{\sum_{s} \sum_{c,c'}\left\{\left[p_{cs}p_{c's}+q_{cs}q_{c's}+ (p_{cs}-q_{cs})(p_{c's}-q_{c's})\right] S_{cs,c's;k}\right\}}\label{iterativeDef}.
\end{eqnarray}
In the results presented in this paper, we set $S_{PQ;0}=S_{PQ}^{(0)}$ as the initial value of the similarity.  This iteration is continued until $\max_{PQ}(|S_{PQ;k+1}-S_{PQ;k}|)\le 0.001$, at which point the algorithm is assumed to have converged.  This relatively high convergence tolerance is due to the computational complexity of the similarity:  there are $\sim N_s\times N_c^2$ (each sector and each pairing of countries for each year) similarities that must be computed, and each requires at on the order of $N_s\times N_c^2$ operations (the number of terms in the sums in Eq. \ref{finalDef}).  This leads to a computational time scaling as $N_s^2N_c^4(\approx 3\times 10^9$ operations for $N_s=35$ and $N_c=41$) to compute one iteration of the of the algorithm.  Convergence to the threshold occurred after $\sim$ 30 minutes on a desktop computer (with the algorithm written in C$++$), and was evaluated on 17 years of data.  

\begin{figure}[H]
\begin{center}
\includegraphics[width=.4\textwidth]{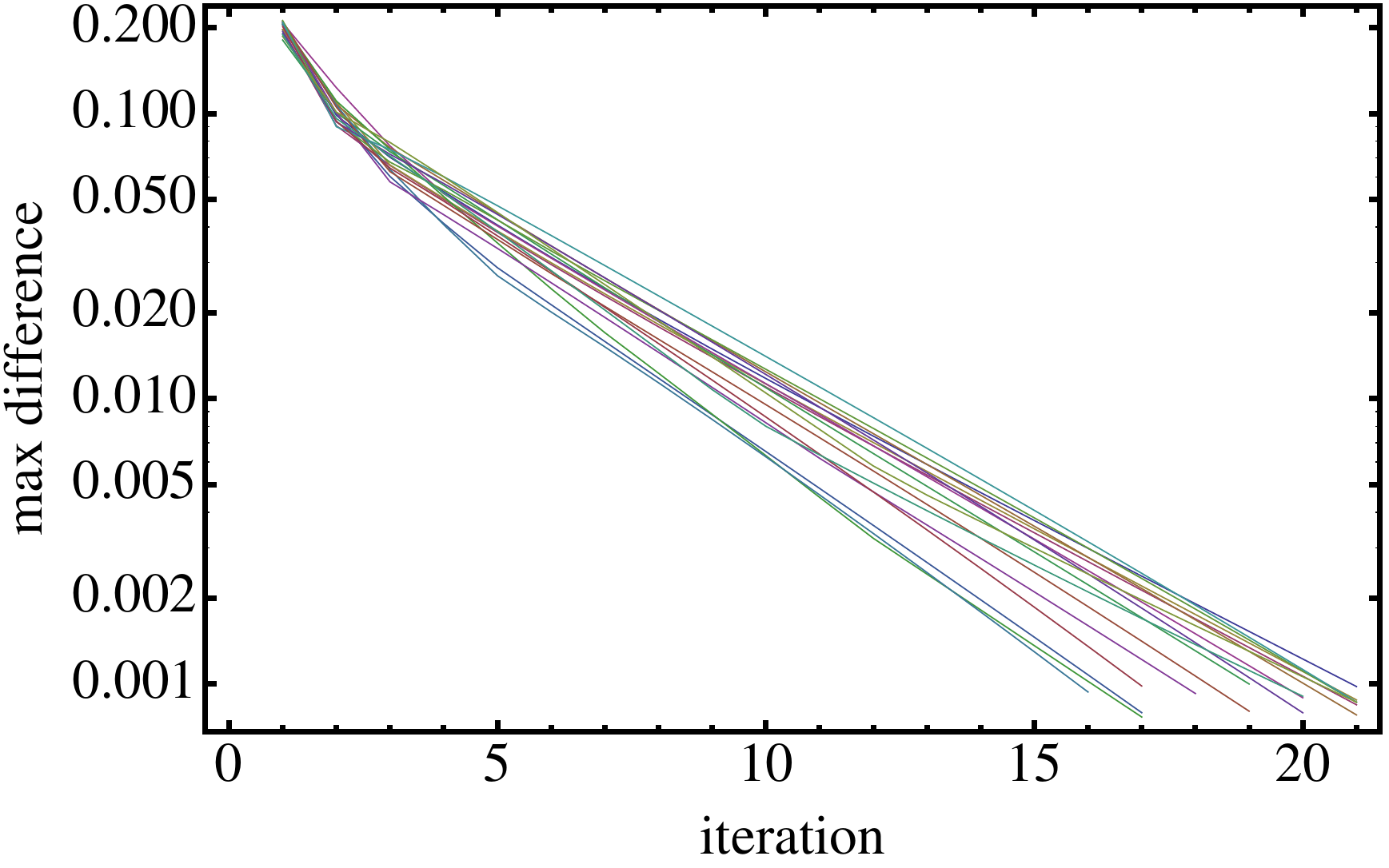}
\caption{The convergence of the algorithm as a function of the iteration.  Each line denotes the maximum difference $\max_{PQ}(|S_{PQ;k+1}-S_{PQ;k}|)$ as a function for the 17 years (1995-2011) on log-linear axes.}
\label{err.fig}
\end{center}
\end{figure}

The method does converge exponentially fast as a function of the iteration (shown in Fig. \ref{err.fig}), and the similarities can be computed after a few hours on a single desktop.  We also compared the values of similarity generated using the initial condition $S_{PQ;0}=S_{PQ}^{(0)}$ with that using the initial condition $S_{PQ;0}=S_{PQ}^{(1)}$ (defined in Eq. \ref{secondDef}), and found that the largest difference between the two measured similarities was on the order of 0.001, the convergence threshold.  This is consistent with the expectation that the algorithm converges to a unique solution.

\subsection{Clustering Countries Based on Similarity}

Blockmodeling tools have been developed in the literature to partition network nodes into clusters according
to structural, automorphic and regular equivalence or other notions of similarity.  The network data are converted into a (dis)similarity matrix, after which some clustering algorithm is applied. In the following we show the clustering of countries after our measure of similarity is applied to compute the distance matrix between countries. We detect some interesting changes over time such as the emergence of a German cluster of upstream interdependencies and the reconfiguration of the relationships among the European countries after the Fifth Enlargement of the European Union in years 2004-2007.

\begin{figure}[H]
 \begin{center}
  \includegraphics[width=\textwidth]{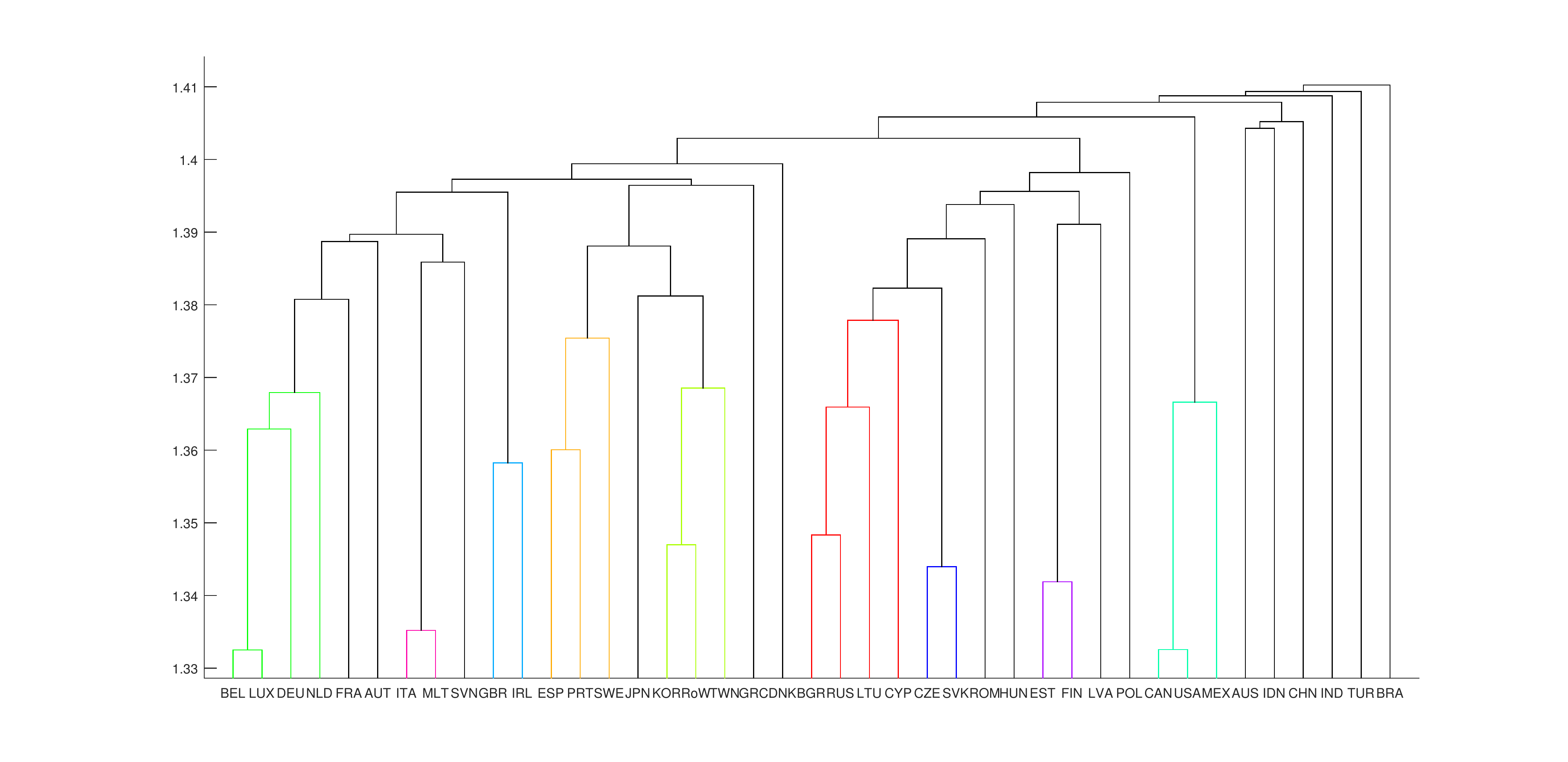}
  \caption{Dendrogram of countries based on unweighted average distance clustering. Distance has been computed as one minus average upstream similarity
	across all sectors in year 1995. Coloring is used to highlight different clusters at a 1.38 cutoff for inter-group dissimilarity. Countries are identified by means of the corresponding	3-characters ISO code.}
  \label{upclust95}
 \end{center}
\end{figure}

\begin{figure}[H]
 \begin{center}
  \includegraphics[width=\textwidth]{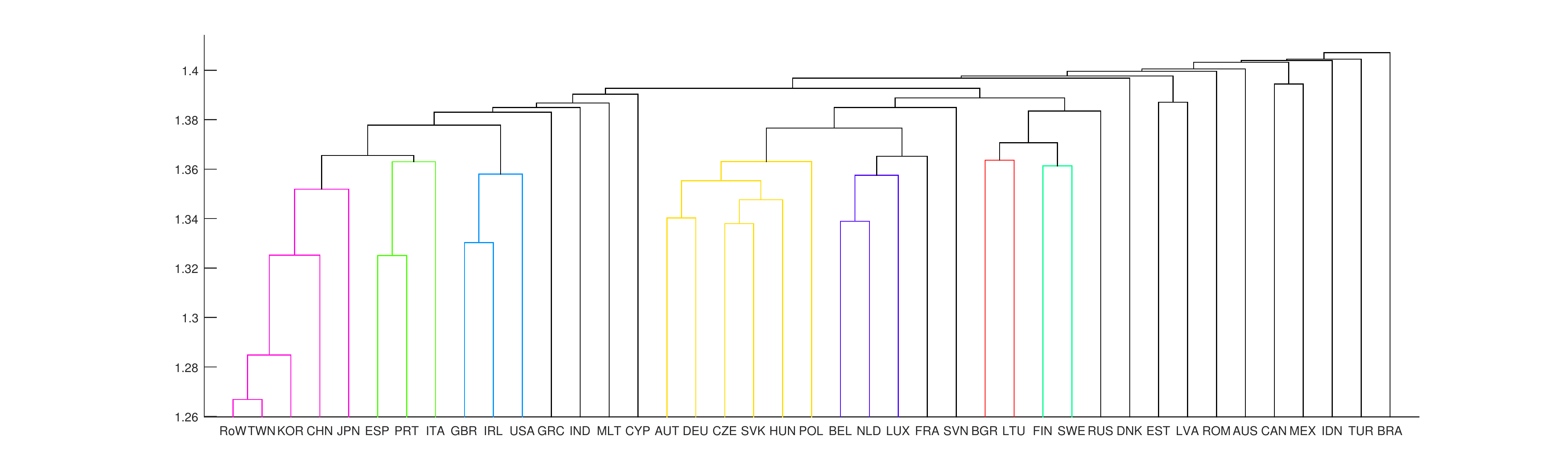}
  \caption{Dendrogram of countries based on unweighted average distance clustering. Distance has been computed as one minus average upstream similarity
	across all sectors in year 2011. Coloring is applied to highlight different clusters at a 1.365 cutoff for inter-group dissimilarity. Countries are identified by means of the corresponding
	3-characters ISO code.}
  \label{upclust11}
 \end{center}
\end{figure}

\begin{figure}[H]
 \begin{center}
  \includegraphics[width=\textwidth]{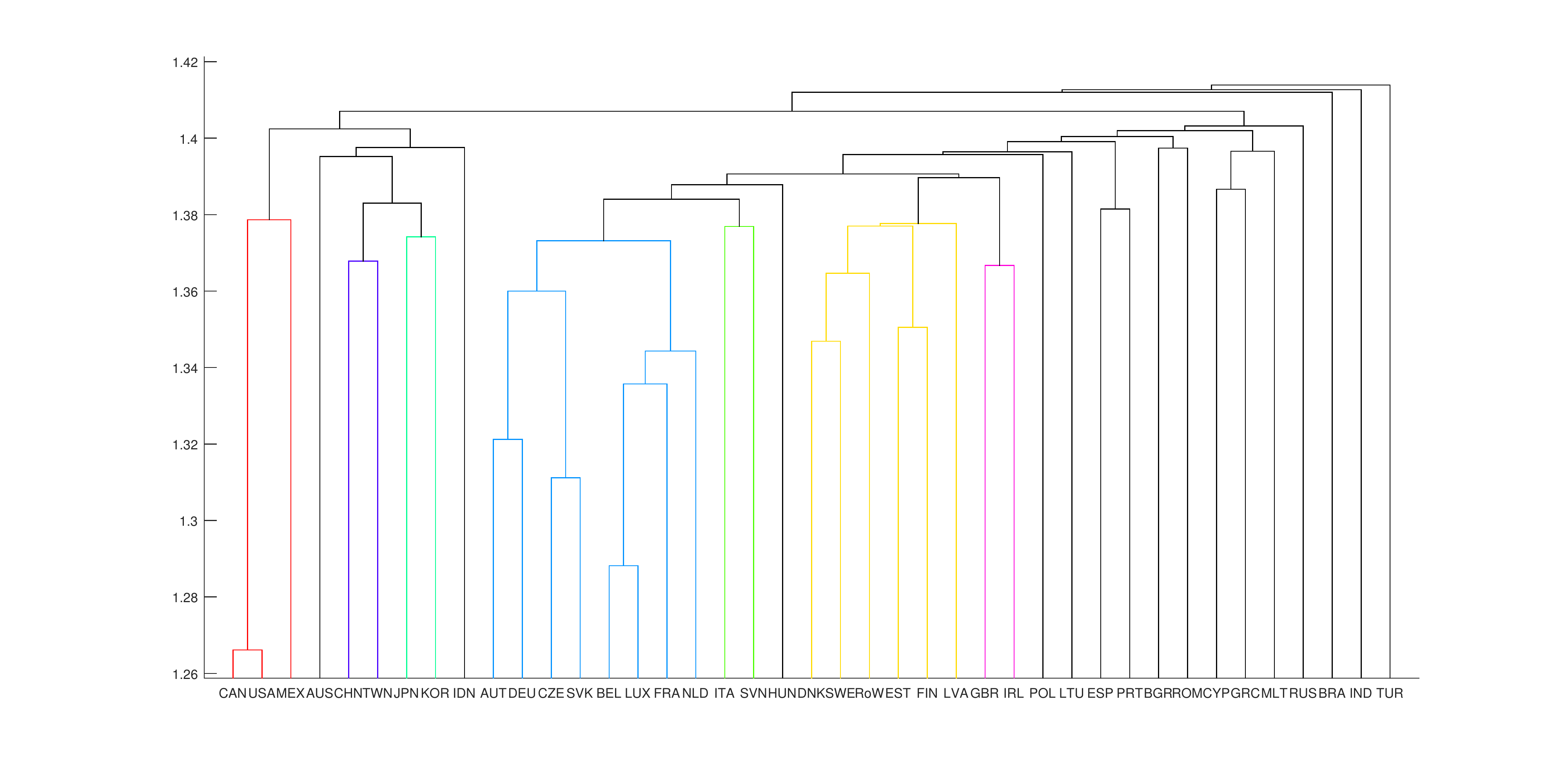}
  \caption{Dendrogram of countries based on unweighted average distance clustering. Distance has been computed as one minus average downstream similarity
	across all sectors in year 1995. Coloring is used to highlight different clusters at a 1.38 cutoff for inter-group dissimilarity. Countries are identified by means of the corresponding
	3-characters ISO code.}
  \label{downclust95}
 \end{center}
\end{figure}

\begin{figure}[H]
 \begin{center}
  \includegraphics[width=\textwidth]{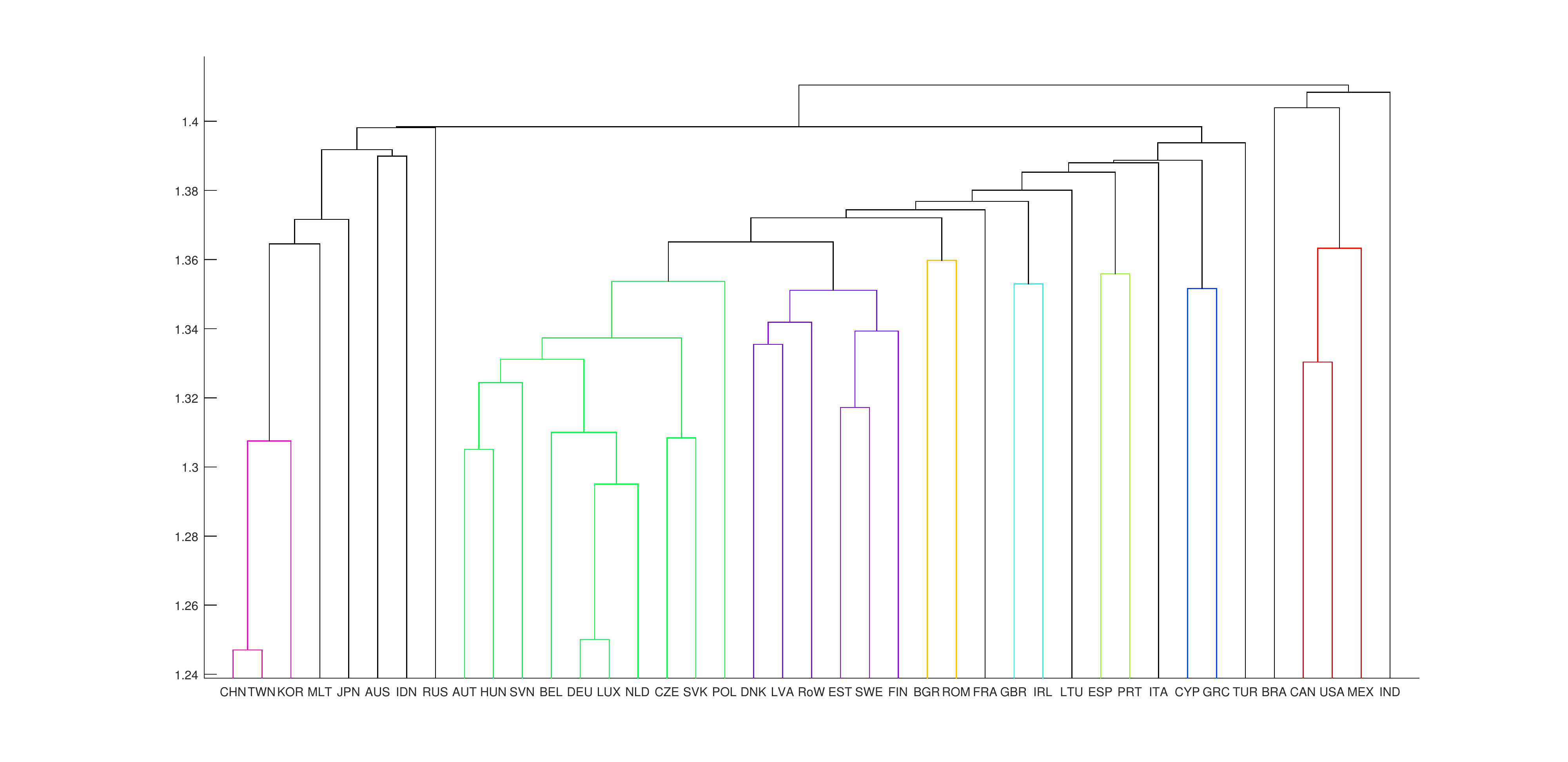}
  \caption{Dendrogram of countries based on unweighted average distance clustering. Distance has been computed as one minus average downstream similarity
	across all sectors in year 2011. Coloring is applied to highlight different clusters at a 1.365 cutoff for inter-group dissimilarity. Countries are identified by means of the corresponding
	3-characters ISO code.}
  \label{downclust11}
 \end{center}
\end{figure}

\section{Acknowledgments}
MR and ZZ acknowledge the funding from the Ministry of Education, Universities and Research (MIUR) - the Basic Research Investment Fund (FIRB) project RBFR12BA3Y and from the ``ViWaN: The Global Virtual Water Network'' project. All authors acknowledge support from the Future Emerging Technologies (FET) projects MULTIPLEX 317532 and SIMPOL 610704 and the National Research Project (PNR) project CRISIS Lab.

\newpage

\bibliographystyle{unsrt}
\bibliography{sim_GVC}

\end{document}